\documentclass[a4paper,12pt]{article}

\usepackage{latexsym}
\usepackage{amsfonts} 
\usepackage[mathscr]{euscript}
\usepackage{amsmath,mathrsfs,bm,amssymb,color}

\title{\textsf{The retarded van der Waals potential - revisited}}
\date{\empty}

\author{
Tadahiro Miyao\footnotemark[1]
  and  Herbert Spohn\footnotemark[2]\\ 
{\it Zentrum Mathematik,}
{\it Technische Universit\"at M\"unchen,}\\ 
{\it  D-85747 Garching, Germany}\\
e-mail:
\footnotemark[1] miyao@ma.tum.de,
\footnotemark[2] spohn@ma.tum.de 
}

\newcommand{\one}{{\mathchoice {\rm 1\mskip-4mu l} {\rm 1\mskip-4mu l}
{\rm 1\mskip-4.5mu l} {\rm 1\mskip-5mu l}}}

\newcommand{\ex}{\mathrm{e}}
\newcommand{\D}{\mathrm{dom}}

\newcommand{\im}{\mathrm{i}}
\newcommand{\Fock}{\mathfrak{F}}

\newcommand{\la}{\langle}
\newcommand{\ra}{\rangle}

\newcommand{\BbbR}{\mathbb{R}}
\newcommand{\BbbN}{\mathbb{N}}

\newcommand{\BbbC}{\mathbb{C}}
\newcommand{\vepsilon}{\varepsilon}
\newcommand{\vphi}{\varphi}

\newcommand{\Hf}{H_{\mathrm{f}}}

\newcommand{\Gr}{\psi}
\newcommand{\TGr}{\tilde{\psi}}
\newcommand{\dm}{\mathrm{d}}
\newcommand{\hphi}{\hat{\vphi}}
\newcommand{\n}{\hat{n}}
\newcommand{\no}{\nonumber \\}

\newcommand{\E}{\mathbb{E}}

\def\Sumoplus{\sideset{}{^{\oplus}_{n\ge 0}}\sum}

\begin{document}

\newtheorem{define}{Definition}[section]
\newtheorem{Thm}[define]{Theorem}
\newtheorem{Prop}[define]{Proposition}
\newtheorem{lemm}[define]{Lemma}
\newtheorem{rem}[define]{Remark}
\newtheorem{assum}{Condition}
\newtheorem{example}{Example}
\newtheorem{coro}[define]{Corollary}

\maketitle

\begin{abstract}
The retarded van-der-Waals potential, as first obtained by Casimir and Polder, is usually  computed  on the basis of nonrelativistic QED. The
hamiltonian describes two infinitely heavy nuclei, charge $e$, separated by a distance $R$ and two spinless electrons, charge $-e$,
 nonrelativistically coupled  to the quantized  radiation field. Casimir and Polder use the dipole  approximation and small coupling to the Maxwell field. We employ here the full
 hamiltonian and determine the asymptotic  
 strength of the leading  $-R^{-7}$ potential, which is   valid for all $e$. Our computation is based on a path integral representation and expands in $1/R$,
 rather than in $e$.
\end{abstract}

\section{Introduction}
\setcounter{equation}{0}
Neutral atoms and molecules interact through the long range, attractive van der Waals potential  which has a decay as $-R^{-6}$ 
for large separation $R$. The quantum origin  of this force was first recognized by London \cite{London}.
Lieb and Thirring \cite{LT} supply a non-perturbative proof valid for very general charge configurations. If one goes beyond
the static Coulomb interaction in quantizing the Maxwell field, then the action is   no longer instantaneous but travels with the speed of light
between 
atoms. In a now  very famous paper \cite{CP} Casimir and Polder establish  that thereby   the effective 
interaction  potential  decays somewhat faster,  namely   as $-R^{-7}$, which is known as  the retarded van der Waals potential. For 
two hydrogen atoms, the cross over between $R^{-6}$ and $R^{-7}$ sets  in at roughly 100 Bohr radii. There are both direct and indirect measurements \cite{Ex} which confirm the theoretical  prediction.

The starting point of Casimir and Polder  is nonrelativistic QED for two atoms separated by  a distance $R$. Within dipole approximation they expand  to fourth order  in the coupling  to the Maxwell field and obtain a prefactor  of $-R^{-7}$ which is proportional to the square of the electric dipole moment
 of a single atom.
  Later on alternative routes and simplified derivations were proposed. For an extensive discussion we refer to the book by Milonni \cite{Milonni},
  see also the monograph by  Margenau and Kestner \cite{MK} and  the lecture notes of Martin and B\"unzli \cite{MB}. Feinberg and Sucher \cite{FS0, FS} reconsider the issue  by employing  a dispersion-theoretic
  approach. Their prefactor  turns out  to be  quadratic in the electric and magnetic dipole moment of a single atom. Somewhat later  Boyer \cite{Boyer}
   rederived the  same prefactor using quantum zero-point energy and semiclassical expressions for the level shifts due to the presence of the atoms.
In our note we stick to nonrelativistic QED, no dipole approximation and no assumption  on  small coupling, and expand in $1/R$. We use the path integral formulation, in which  the subtraction of the ground state energy at $R=\infty$ is particularly transparent.
As in previous studies  the strength of the retarded van der Waals potential is quadratic  in the electric and magnetic dipole moments, but with  modified  coefficients
as compared   to \cite{FS, Boyer}.

In mathematical physics there has been  a revived interest in nonrelativistic QED \cite{Spohn}. It is conceivable that some parts of the argument can be elevated to a rigorous proof. In our paper we mostly ignore this line of research, but will provide   a more detailed  discussion in the conclusions.

\section{Hamiltonian and van der Waals potential}
\setcounter{equation}{0}

We consider a single hydrogen atom with an infinitely heavy nucleus located at the origin. The nucleus has charge $e$, $e>0$,
the electron has charge $-e$. We will use units in which $\hbar=1, c=1$, and the bare mass of the electron $m=1$. 
In Section \ref{DL} we will restore the proper physical units.
 Let $x,p$ be position and momentum of the electron. Then the nonrelativistic QED hamiltonian for this system reads 
\begin{align}
H=\frac{1}{2}\big(p-eA(x)\big)^2-e^2V(x)+\Hf.\label{1PL}
\end{align}
For $H$ to make sense the electron is assumed to have a prescribed  charge distribution $\vphi$ with the following properties: $\vphi$ is  normalized,
$\int \mathrm{d}x\, \vphi(x)=1$, rotation invariant,
$ \vphi(x)=\vphi_{\mathrm{rad}}(|x|)$,  of rapid decrease, and its Fourier transform,  $\hat{\vphi}$,  is   real. Then
$V$ is the smeared Coulomb potential
\begin{align}
V(x)=\int\mathrm{d}k\, |\hat{\vphi}(k)|^2|k|^{-2} \ex^{-\im k\cdot x}.
\end{align}
$A(x)$ is the quantized  vector potential and $\Hf$  is the field energy. These are defined through a two-component Bose field 
$a(k, \lambda), k\in \BbbR^3, \lambda=1,2,$ with commutation relation
\begin{align}
[a(k, \lambda), a(k', \lambda')^*]= \delta_{\lambda\lambda'}\delta(k-k').
\end{align}
Explicitly
\begin{align}
\Hf=\sum_{\lambda=1,2}\int_{\BbbR^3}\mathrm{d}k\, \omega(k)a(k, \lambda)^*a(k, \lambda)
\end{align}
with dispersion  relation
\begin{align}
\omega(k)=|k|
\end{align}
and 
\begin{align}
A(x)=\sum_{\lambda=1,2}\int_{\BbbR^3}\mathrm{d}k\, \frac{\hat{\vphi}(k)}{\sqrt{2\omega(k)}}\vepsilon(k, \lambda)\big(\ex^{\im k\cdot x}a(k, \lambda)+\ex^{-\im k\cdot x}a(k, \lambda)^*\big)
\end{align}
with the standard dreibein $\vepsilon(k,1), \vepsilon(k, 2), \hat{k}=k/|k|$. Thus the Hilbert space for $H$ is 
\begin{align}
\mathcal{H}=L^2(\BbbR^3_x)\otimes \mathfrak{F},
\end{align} 
where $\mathfrak{F}$ is the bosonic Fock space over $
L^2(\BbbR^3)\otimes \BbbC^2$.
From the quantization  of the classical system of charges coupled to
the Maxwell field it follows that for the smearing of $A(x)$ and of
$V$ the same charge distribution has to be used. We refer to
\cite{Spohn} for details.
As proved by Griesemer, Lieb,  and Loss \cite{BFS, LL, GLL}, $H$ has a unique
ground state, denoted here  by $\Gr$, with ground state energy $E$, $H\Gr=E\Gr$.

The asymptotic strength $\kappa $ of the van der Waals potential  depends on the properties of a single hydrogen atom only through its electric and magnetic
dipole moment, $\alpha_{\mathrm{E}}$ and $\alpha_{\mathrm{M}}$. They
are defined  through the energy, $W$,  of our   system for weak external uniform electric and  magnetic fields according to 
\begin{align}
W=-\frac{1}{2} \alpha_{\mathrm{E}} E_{\mathrm{ex}}^2-\frac{1}{2} \alpha_{\mathrm{M}} B_{\mathrm{ex}}^2.
\end{align}
To say,  $H$ is perturbed by $e E_{\mathrm{ex}}\cdot x$ and the vector potential  is perturbed by $\frac{1}{2} B_{\mathrm{ex}}\wedge x$.
Then by second order perturbation theory  it follows that 
\begin{align}  
\alpha_{\mathrm{E}}&=2 \big( \tfrac{1}{3} \la \Gr, x\cdot (H-E)^{-1} x\Gr\ra\big)e^2,
\no
\alpha_{\mathrm{M}}&= -\frac{1}{4} \big( \tfrac{1}{3} \la \Gr, x^2\Gr\ra\big)e^2.
\end{align} 
As a convention, $\la\cdot, \cdot \ra$ denotes always the inner
product on the respective Hilbert space.

To investigate  the van der Waals potential we consider  two hydrogen atoms, one
located at $0$ and the other at $r=(0,0,R), R\ge 0$. It will be  convenient to
define the position of the second electron relative to $r$. Then $x_1,
x_2+r$ are positions  and $p_1, p_2$ the momenta of the two
electrons. The two-electron hamiltonian reads
\begin{align}
H_R=&\frac{1}{2}\big(p_1-eA(x_1)\big)^2-e^2V(x_1)+\frac{1}{2}\big(p_2-eA(x_2+r)\big)^2-e^2V(x_2)\nonumber \\
&+\Hf+e^2 V_R(x_1, x_2)\label{FullHami}
\end{align}
with the interaction potential
\begin{align}
V_R(x_1, x_2)&= -V(x_1-r)-V(x_2+r)+V(r)+V(r+x_2-x_1)\nonumber \\
&=\int_{\BbbR^3}\mathrm{d}k\, |\hat{\vphi}(k)|^2\ex^{\im k\cdot r}
|k|^{-2}
(1-\ex^{-\im k\cdot x_1})(1-\ex^{\im k\cdot x_2}).
\end{align}
$H_R$  acts on the Hilbert space $L^2(\BbbR^3_{x_1})\otimes
L^2(\BbbR^3_{x_2})\otimes \mathfrak{F}$. $H_R$ has a unique ground
state with energy $E(r)$.

In the Born-Oppenheimer approximation $E(r)$ is the effective
potential  between the two neutral hydrogen atoms in their ground state. Thus the issue at
hand is to investigate $E(r)$ for large $R$. By  rotation invariance $E(r)$ depends only on $|r|=R$ and 
we also write $E(r)=E(R)$. 
 For $R\to \infty$ the two atoms become independent and  one
can show that $E(R)$ converges to $2E$. The Casimir-Polder result  is
that, for small $e$,
\begin{align}
\lim_{R\to \infty}R^7(E(R)-2E)=-\frac{23}{4\pi}\big(\frac{1}{2\pi}\big)^2
\big(\frac{1}{2}\alpha_{\mathrm{E, at}}\big)^2 .\label{CP}
\end{align} 
The factor $(1/2\pi)^2$ results from our use of the Lorentz-Heaviside
units.
We remark that Casimir and Polder omit in their definition of $\alpha_{\mathrm{E}}$ the factor $2$, which accounts for the extra $1/2$.
 $\alpha_{\mathrm{E, at}}$ is the dipole moment of a decoupled hydrogen atom. 
It is defined through
\begin{align}
\alpha_{\mathrm{E}, \mathrm{at}}= 2 \big(\tfrac{1}{3} \la\psi_{\mathrm{at}},
x\cdot (H_{\mathrm{at}}-E_{\mathrm{at}})^{-1} x\psi_{\mathrm{at}}\ra \big)e^2
\label{Prefactor}
\end{align}
with 
\begin{align}
H_{\mathrm{at}}=\frac{1}{2}p^2-e^2 V(x)
\end{align}
and $\psi_{\mathrm{at}}$   the ground state of the hydrogen atom,
$H_{\mathrm{at}}\psi_{\mathrm{at}}=E_{\mathrm{at}}\psi_{\mathrm{at}}$.
 
 In our  set-up, the natural dimensionless coupling constant is the Sommerfeld fine-structure constant 
 \begin{align}
 \alpha=\frac{e^2}{4\pi \hbar c}.
 \end{align} 
 The energy unit is set by the ionization energy of the hydrogen atom,
 which is  $\alpha^2 m c^2$ and  the length unit is the Bohr radius
 $r_{\mathrm{B}}=\hbar/\alpha m c$. Anticipating  a decay as $R^{-7}$,
 the dimensionless coupling strength, $\kappa$,  is defined through
 \begin{align}
 E(R)-2E =-\kappa(\alpha, \lambda_{\mathrm{c}}\Lambda) \alpha m c^2 (R/r_{\mathrm{B}})^{-7} 
 \end{align}
 valid for large $R$. $\kappa$ depends on $\alpha$ and on the ultraviolet cutoff $\Lambda$ in units of the inverse Compton wave length $\lambda_{\mathrm{c}}=\hbar / mc$. For this  interpretation the form factor $\hat{\vphi}$ is chosen as  $\hat{\vphi}(k)=\hat{\vphi}_1(\Lambda^{-1}k)$, 
 where $\hat{\vphi}_1$ rapidly interpolates between  $\hat{\vphi}_1=(2\pi)^{-3/2}$ for $|k|\le 1-\delta$ and $\hat{\vphi}_1=0$ for $|k|\ge 1$.

The goal of our note is to obtain an exact expression for the strength $\kappa$. Readers not so much interested in the details  of the computation
 may skip ahead  to Section \ref{DL} where the result  is discussed.

\section{Path integration}\setcounter{equation}{0}

As noted by Feynman \cite{Feynman}, in the functional integral
representation
 of $\ex^{-tH_R}, t\ge 0$, the interaction with the radiation field
 is linear in $A$. Therefore one can carry out the Gaussian
 integration over the fluctuating photon field. This form will be particularly
 convenient for the Born-Oppenheimer energy $E(R)-2E$. After such a detour we
 will return to operators. Our notation is formal, but rigorous
 versions
are available \cite{Hiroshima1, Hiroshima2}.

We denote by $q_j(t)\in \BbbR^3$ the path of electron $j$. In case of
a single electron we omit the index $j$. For the ground state energy
of  the two-electron system  one obtains 
\begin{align}
E(r)=&-\lim_{T\to \infty}\frac{1}{2T}\log
\int\Big[\Pi\mathrm{d}q_1(\cdot)\Big]\int\Big[\Pi\mathrm{d}q_2(\cdot)\Big]
\nonumber \\
&\exp\Big[-\int_{-T}^T\mathrm{d}t\Big(\sum_{j=1,2}\big(\tfrac{1}{2}\dot{q}_j(t)^2-e^2V(q_j(t))
\big)
+e^2 V_R(q_1(t), q_2(t))
\Big)\nonumber \\
&-\sum_{j=1,2}\tfrac{1}{2}e^2\int_{-T}^T\mathrm{d}s\int_{-T}^T\mathrm{d}t\,
\dot{q}_j(s)\cdot W_0\big(q_j(s)-q_j(t), s-t\big)\dot{q}_j(t)\nonumber
\\
&- e^2\int_{-T}^T\mathrm{d}s\int_{-T}^T\mathrm{d}t\,
\dot{q}_1 (s)\cdot W_R\big(q_1(s)-q_2(t), s-t\big)\dot{q}_2(t)
\Big].
\end{align}
Here $\int\big[\Pi\mathrm{d}q_j(\cdot)\big]$ is the ``sum  over all
paths"  and $W_R$ is the photon propagator,
 \begin{align}
W_R(q,t)=\int_{\BbbR^3}\mathrm{d}k\,
|\hat{\vphi}(k)|^2\frac{1}{2\omega(k)}
(\one-| \hat{k}\ra\la\hat{k}|)\, \ex^{-\omega(k)|t|}\ex^{\im k\cdot
  r}\ex^{-\im k\cdot q}\label{Propagator}
\end{align}
as a  $3\times 3$ matrix.
Here $\one $ is the unit matrix and   $|\hat{k}\ra\la \hat{k}|$ the orthogonal
projection onto $\hat{k}$, $\hat{k}=k/|k|$.

Correspondingly for a single electron
\begin{align}
E=&-\lim_{T\to \infty}\frac{1}{2T}\log \int
\Big[\Pi \mathrm{d}q(\cdot)\Big]
\exp\Big[
-\int_{-T}^T \mathrm{d}t\, \big(\tfrac{1}{2}\dot{q}(t)^2-e^2 V(q(t))\big)
\no
&-\tfrac{1}{2}e^2\int_{-T}^T\mathrm{d}s\int_{-T}^T\mathrm{d}t\,
\dot{q}(s)\cdot W_0\big(q(s)-q(t), s-t\big)\dot{q}(t)
\Big].\label{SingleEnergy}
\end{align}
Therefore
\begin{align}
E(R)-2E=&-\lim_{T\to \infty}\frac{1}{2T}\log \E\times \E \Big[ 
\exp\Big[
-e^2\int_{-T}^T\mathrm{d}t\, V_R(q_1(t), q_2(t))
\no
&-e^2\int_{-T}^T\mathrm{d}s\int_{-T}^T\mathrm{d}t\,
\dot{q}_1(s)\cdot W_R\big(q_1(s)-q_2(t), s-t\big)\dot{q}_2(t)
\Big]
\Big].\label{PathBindingEr}
\end{align}
Here $q_1(t)$ and $q_2(t)$  are two independent copies  of the ground state process $q(t)$ with a  path measure as written in (\ref{SingleEnergy}).
$q(t)$ is stationary and the
distribution of $q(t)$ at a fixed time is the electronic density
computed from  the ground state    $\Gr$ of $H$.
The average with respect to the ground state process is denoted by $\E[\cdot]$ and the average over the two independent copies by $\E\times \E[\cdot]$

We note that in the  expression (\ref{PathBindingEr}) the only $R$
dependence sits in $V_R$ and $W_R$, which in a certain sense are
small. Thus it is  natural to use the cumulant expansion.
Denoting the exponent by $X_T(R)$, one arrives at 
\begin{align}
-\frac{1}{2T}\log \E\times \E[ \ex^{- e^2 X_T(R)}] 
&=-\frac{1}{2T}\big(
e^2 C_1(R,T)+\frac{1}{2}e^4 C_2(R, T)+\cdots\label{2ndCum}
\big),\nonumber\\
C_1(R, T)&= -\E\times \E[X_T(R)],\nonumber \\[1ex]
C_2(R,T)&=\E\times \E [X_T(R)^2] -\E\times \E [X_T(R)]^2.
\end{align}
As we will see below $\E\times \E [ X_T(R)^2]/2T=\mathcal{O}(R^{-6})$, not using the cancellation between  $V_R$  and $W_R$ terms.
Since we are heading for a decay as  $R^{-7}$    for large $R$,  it should
 suffice to stop the expansion at the second cumulant $C_2(R, T)$.
{\it  This will be our main assumption.}
$\E\times \E[ X_T(R)]/2T$ is exponentially small. The large $R$ behavior of the
second cumulant will be investigated  in detail in the following section.

We note that the expectation  under the $k$-integral always factorizes with respect to
$q_1, q_2$. Using their independence we will need to compute 
only a few expectations for the ground state process of a single electron. They are listed
now for  later convenience and proved in Appendix \ref{(i)to(v)}. We set $\vepsilon=\vepsilon(k, \lambda)$,  
$\vepsilon_j=\vepsilon(k_j, \lambda_j)$ for either $\lambda_j=1$
or $\lambda_j=2$ and $\theta(t)$ the step function, $\theta(t)=-1$ for $t\le 0$, $\theta(t)=1$ for $t> 0$. 
\begin{itemize}
\item[{\rm (i)}] $\displaystyle  \E[ \ex^{\im k\cdot q(t)}]=\la \Gr,
  \ex^{\im k\cdot x}\Gr\ra$,
\item[{\rm (ii)}] $\displaystyle \E[ \ex^{\im k\cdot q(t)}(\vepsilon(k,
  \lambda)\cdot \dot{q}(t))]=-\la \Gr, \ex^{\im k\cdot
  x}(H-E)(\vepsilon(k, \lambda)\cdot x)\Gr\ra=0 ,$
\item[{\rm (iii)}] $\displaystyle \E[ \ex^{-\im k_1\cdot q(s)}\ex^{\im
  k_2\cdot q(t)}]=\la \Gr, \ex^{-\im k_1\cdot
  x}\ex^{-|s-t|(H-E)}\ex^{\im k_2\cdot x}\Gr\ra$,
\item[{\rm (iv)}] $ \displaystyle  \E [(\vepsilon_1\cdot
  \dot{q}(s))\ex^{-\im k_1\cdot q(s)}(1-\ex^{-\im k_2\cdot q(t)})]\vspace{3mm}
\\=
\theta(t-s)\la \Gr, (\vepsilon_1\cdot x)(H-E)\ex^{-\im
  k_1\cdot x} \ex^{-|s-t|(H-E)}(1-\ex^{-\im k_2\cdot x})\Gr\ra,
$
 \item[{\rm (v)}]$\displaystyle 
 \E [(\vepsilon_1\cdot \dot{q}(s))\ex^{-\im k_1\cdot q(s)}\ex^{-\im
  k_2\cdot q(t)}(\vepsilon_2\cdot \dot{q}(t))] \vspace{3mm}
  \\
 =-\la\Gr,  (\vepsilon_1\cdot x )(H-E)\ex^{-\im k_1\cdot
  x}\ex^{-|s-t|(H-E)}\ex^{-\im k_2\cdot x} (H-E)(\vepsilon_2\cdot
x)\Gr\ra\vspace{3mm}\\
 +\delta(t-s)(\vepsilon_1\cdot \vepsilon_2)\la \Gr, \ex^{-\im k_1\cdot x}
\ex^{-\im k_2\cdot x}\Gr\ra. \label{(v)}
$
\end{itemize}
 In (v) the second term arises because  locally $q(t)$ is
like a standard Brownian motion  for  which  $\mathrm{d}q(s)\otimes
\mathrm{d}q(t)=\one \delta(s-t)\mathrm{d}s\mathrm{d}t.
$

The first cumulant can be dealt with immediately.
Using (ii) in the above list one obtains 
\begin{align}
\E\times \E\big[ \dot{q}_1(s)\cdot W_R\big(q_1(s)-q_2(t), s-t\big)\dot{q}_2(t)\big]=0.
\end{align} 
For the potential term it holds
\begin{align}
&\E\times \E[ V_R(q_1(t), q_2(t))]
\no
&=\int_{\BbbR^3}\mathrm{d}k\,
|\hat{\vphi}(k)|^2\omega(k)^{-2}\ex^{\im k\cdot r} 
\la \Gr, (1-\ex^{-\im k\cdot x})\Gr\ra\la \Gr, (1-\ex^{\im k\cdot
  x})\Gr\ra
  \no
&=\int_{\BbbR^3}\mathrm{d}x\int_{\BbbR^3}\mathrm{d}y\big((\varrho-\delta)*\vphi\big)(x)
(4\pi |x+r-y|)^{-1}\big((\varrho-\delta)*\vphi\big)(y).
\end{align}
Here $\varrho(x)$ is the electron density for $\Gr$, which is known  to
have an exponential decay \cite{Gr}. $\delta$ is the Dirac delta
and $*$ denotes convolution. Since $\int_{\BbbR^3}\mathrm{d}x\,
(\varrho(x)-\delta(x))=0$ and since $\varrho$ decays rapidly, by
Newton's theorem  it holds that there are suitable constants $c_1,
c_2$ such that 
\begin{align}
\big|\E\times \E[ V_R(q_1(t), q_2(t))]\big| \le c_1 \ex^{-c_2R}.
\end{align}

\section{The second cumulant}\setcounter{equation}{0}
\label{Computation}

The second cumulant consists of three terms. They are all
proportional to $2T$ by the stationarity of $q_1(t), q_2(t)$.
We could take the limit $T\to \infty$ first, but the symmetric version
is more convenient. All expectations are written in terms of  $H$ and
its ground state $\Gr$. For notational symplicity we replace  $H-E$ by
$H$,
 hence $H\Gr=0$. For inverses as $\la \phi_1,
 H^{-n}\phi_2\ra$
we make sure that either $\la \phi_1, \Gr\ra=0$ or $\la
\phi_2, \Gr\ra=0$. Note that the ground state is nondegenerate \cite{Hiroshima1}.
But $\la \phi_1, H^{-n}\phi_2\ra$ could still be
infinite.
If $H$ is replaced by $H_{\mathrm{at}}$, then $H_{\mathrm{at}}$ has a  spectral gap  and therefore  an inverse on the
orthogonal  complement of $\psi_{\mathrm{at}}$.

We set
\begin{align}
C_2(R, T)=I_{VV}+2I_{VW}+I_{WW}\label{Dec2ndCum}
\end{align}
and compute each term separately.
The second step is a partial time-integration through which one can
understand how the $R^{-6}$ decay from the interaction potential is
canceled.
 In a final step we collect terms according to their number of
 time-integrations and discuss their $R$-dependence.

\subsection{Expectations}
  a) $I_{VV}$. Setting $\hat{\vphi}(k_j)=\hat{\vphi}_j,
  \omega(k_j)=\omega_j$, 
we have 
\begin{align}
I_{VV}=&\int\mathrm{d}t_1\mathrm{d}t_2\int\mathrm{d}k_1\mathrm{d}k_2\, 
|\hat{\vphi}_1|^2|\hat{\vphi}_2|^2(|k_1|^2|k_2|^2)^{-1}\ex^{\im
  (k_1+k_2)\cdot r}
  \nonumber\\
&\times \Big\{
\big|
\E[(1-\ex^{-\im k_1\cdot q_1(t_1)})(1-\ex^{-\im k_2\cdot q_1(t_2)})]
\big|^2
\nonumber\\
&-\big|
\E[ (1-\ex^{-\im k_1\cdot q_1(t_1)})]\E[(1-\ex^{-\im k_2\cdot q_1(t_2)})]
\big|^2
\Big\}
\no
=&\int\mathrm{d}t_1\mathrm{d}t_2\int\mathrm{d}k_1\mathrm{d}k_2\, 
|\hat{\vphi}_1|^2|\hat{\vphi}_2|^2(|k_1|^2 |k_2|^2)^{-1}\ex^{\im
  (k_1+k_2)\cdot r}
  \nonumber\\
&\times \Big\{
\big|
\la\Gr,  (1-\ex^{-\im k_1\cdot x})\ex^{-|t_1-t_2|H}(1-\ex^{-\im k_2\cdot x})\Gr\ra
\big|^2
\nonumber\\
&-\big|
\la\Gr, (1-\ex^{-\im k_1\cdot x})\Gr\ra\la\Gr, (1-\ex^{-\im k_2\cdot x})\Gr\ra
\big|^2
\Big\}. \label{IVV}
\end{align} 

Note that  the integrand in (\ref{IVV}) decays to zero, since 
\begin{align}
\ex^{-|t_1-t_2|H}\to |\Gr \ra\la \Gr |
\end{align}
as $|t_1-t_2|\to \infty$.\medskip\\\medskip\\
b) $I_{VW}$. Setting $\vepsilon_j=\vepsilon(k_j, \lambda_j)$ and noting 
\begin{align}
\sum_{\lambda=1,2}| \vepsilon(k, \lambda)\ra\la \vepsilon(k, \lambda)|=\one -| \hat{k}\ra \la \hat{k}|, \label{Polarization1}
\end{align}
 one has 
\begin{align}
I_{VW}=&\int\dm t_1\dm t_2 \dm t_3\sum_{\lambda_1=1,2} \int \dm k_1\dm k_2 |\hphi_1|^2|\hphi_2|^2 (2\omega_1 |k_2|^2)^{-1}
\ex^{\im (k_1+k_2)\cdot r} \ex^{-\omega_1|t_1-t_2|}
\nonumber\\ 
&\times \E[ (\vepsilon_1\cdot \dot{q}_1(t_1))\ex^{-\im k_1\cdot q_1(t_1)}(1-\ex^{-\im k_2\cdot q_1(t_3)})]
\nonumber\\
&\times  \E[  (\vepsilon_1\cdot \dot{q}_2(t_2))\ex^{\im k_1\cdot q_2(t_2)}(1-\ex^{\im k_2\cdot q_2(t_3)})]
\no
=&\int\dm t_1\dm t_2 \dm t_3\sum_{\lambda_1=1,2} \int \dm k_1\dm k_2 |\hphi_1|^2|\hphi_2|^2 (2\omega_1 |k_2|^2)^{-1}
\ex^{\im (k_1+k_2)\cdot r} \ex^{-\omega_1|t_1-t_2|}
\no
&\times \theta(t_1-t_3)\la\Gr,  (\vepsilon_1\cdot x) H \ex^{-\im k_1\cdot x}\ex^{-|t_1-t_3|H}(1-\ex^{-\im k_2\cdot x})\Gr\ra
\no
&\times \theta(t_2-t_3)\la\Gr,  (\vepsilon_1\cdot x) H \ex^{\im k_1\cdot x}\ex^{-|t_2-t_3|H}(1-\ex^{\im k_2\cdot x})\Gr\ra.
 \end{align}
As proved in Appendix \ref{GrEx}, it holds
\begin{align}
\la\Gr,  (\vepsilon_1\cdot x)H \ex^{-\im k_1\cdot x}\Gr\ra=\im \la \Gr, (\vepsilon_1\cdot (p-eA(x)))\ex^{-\im k_1\cdot x}\Gr\ra=0.
\end{align}
Thus no truncation  of the expectation is needed.
\medskip\\
\medskip\\
c) $I_{WW}$. One has 
\begin{align}
I_{WW}
=&\int \dm t_1\dm t_2\dm t_3\dm t_4 \sum_{\lambda_1, \lambda_2}\int \dm k_1\dm k_2 |\hphi_1|^2 |\hphi_2|^2(2\omega_12\omega_2)^{-1}
\ex^{\im (k_1+k_2)\cdot r} 
\no
&\times \ex^{-\omega_1|t_1-t_2|}\ex^{-\omega_2|t_3-t_4|}
 \E\big[ (\vepsilon_1 \cdot \dot{q}_1(t_1)) \ex^{-\im k_1\cdot q_1(t_1)}(\vepsilon_2 \cdot \dot{q}_1(t_3))\ex^{-\im k_2\cdot q_1(t_3)}\big]
 \no
 &\times \E\big[ (\vepsilon_1 \cdot \dot{q}_2(t_2)) \ex^{\im k_1\cdot q_2(t_2)}(\vepsilon_2 \cdot \dot{q}_2(t_4))\ex^{\im k_2\cdot q_2(t_4)}\big]
 \no
 =&\int \dm t_1\dm t_2\dm t_3\dm t_4 \sum_{\lambda_1, \lambda_2}\int \dm k_1\dm k_2 |\hphi_1|^2 |\hphi_2|^2(2\omega_12\omega_2)^{-1}
\ex^{\im (k_1+k_2)\cdot r} 
\no
&\times \ex^{-\omega_1|t_1-t_2|}\ex^{-\omega_2|t_3-t_4|}
\no
&\times\big(
 -\la\Gr,  (\vepsilon_1 \cdot x) H  \ex^{-\im k_1\cdot x} \ex^{-|t_1-t_3|H}\ex^{-\im k_2\cdot x }H  (\vepsilon_2 \cdot x)\Gr\ra
 \no
 &+ \delta(t_1-t_3)(\vepsilon_1\cdot \vepsilon_2) \la \Gr, \ex^{-\im (k_1+k_2)\cdot x}\Gr\ra
 \big)
 \no
 &\times \big(
- \la\Gr,  (\vepsilon_1 \cdot x) H  \ex^{\im k_1\cdot x} \ex^{-|t_2-t_4|H}\ex^{\im k_2\cdot x }H  (\vepsilon_2 \cdot x)\Gr\ra
\no
 &+ \delta(t_2-t_4)(\vepsilon_1\cdot \vepsilon_2) \la \Gr, \ex^{\im (k_1+k_2)\cdot x}\Gr\ra
 \big).
  \end{align}

  \subsection{Partial time integration}
  The next step is a partial time integration for $I_{VW} $ and $I_{WW}$. For the integrand of $I_{VW}$ we use the identity
  \begin{align}
 & \theta(s-t) \la \Gr, (\vepsilon_1\cdot x) H \ex^{-\im k_1\cdot x} \ex^{-|s-t|H} (1-\ex^{-\im k_2\cdot x})\Gr\ra
  \no
  &=-\frac{\partial }{\partial s} \la \Gr, ( \vepsilon_1\cdot x) H \ex^{-\im k_1\cdot x} H^{-1} \ex^{-|s-t|H} (1-\ex^{-\im k_2 \cdot x})\Gr\ra
  \end{align}
and for the integrand of $I_{WW}$ the identity
\begin{align}
&-\la \Gr, (\vepsilon_1\cdot x) H \ex^{-\im k_1\cdot x }\ex^{-|s-t|H} \ex^{-\im k_2\cdot x} H (\vepsilon_2\cdot x)\Gr\ra
\no
&=\frac{\partial}{\partial s}\frac{\partial}{\partial t} \la \Gr, (\vepsilon_1\cdot x) H \ex^{-\im k_1\cdot x }H^{-2} \ex^{-|s-t|H} \ex^{-\im k_2\cdot x} H (\vepsilon_2\cdot x)\Gr\ra
\no
& -2 \delta(s-t)\la \Gr, (\vepsilon_1\cdot x) H \ex^{-\im k_1\cdot x}H^{-1} \ex^{-\im k_2\cdot x}H (\vepsilon_2 \cdot x)\Gr\ra.
\end{align}
We insert these expectations in $I_{VW}, I_{WW}$, integrate partially in time, and use
\begin{align}
\frac{\partial}{\partial s}\frac{\partial}{\partial t}\ex^{-\omega|s-t|}=-\omega^2\ex^{-\omega|s-t|}+2\omega\delta(s-t).
\end{align}
The boundary terms vanish. For $I_{VW} $ one obtains
\begin{align}
I_{VW}=&\int\dm t_1\dm t_2\dm t_3 \sum_{\lambda_1}\int\dm k_1\dm k_2 |\hphi_1|^2|\hphi_2|^2 (2\omega_1|k_2|^2)^{-1} \ex^{\im (k_1+k_2)\cdot r}
\no
&\times \big(-\omega_1^2\ex^{-\omega_1|t_1-t_2|}+2\omega_1 \delta(t_1-t_2)\big)\no
&\times \la \Gr, (\vepsilon_1\cdot x) H \ex^{-\im k_1\cdot x}H^{-1}\ex^{-|t_1-t_3|H}(1-\ex^{-\im k_2\cdot x})\Gr\ra
\no
&\times \la \Gr, (\vepsilon_1\cdot x) H \ex^{\im k_1\cdot x} H^{-1} \ex^{-|t_2-t_3|H} (1-\ex^{\im k_2\cdot x})\Gr\ra.
\end{align}
Finally  $I_{WW}$  is given by
\begin{align}
I_{WW}=&\int \dm t_1\dm t_2\dm t_3\dm t_4 \sum_{\lambda_1, \lambda_2}\int \dm k_1\dm k_2 |\hphi_1|^2 |\hphi_2|^2(2\omega_12\omega_2)^{-1}
\ex^{\im (k_1+k_2)\cdot r} 
\no
&\times\Big\{ \big(-\omega_1^2 \ex^{-\omega_1|t_1-t_2|}+2\omega_1 \delta(t_1-t_2)\big)\big(-\omega_2^2 \ex^{-\omega_2|t_3-t_4|}+2\omega_2 \delta(t_3-t_4)\big)
\no
&
 \times \la\Gr,  (\vepsilon_1 \cdot x) H  \ex^{-\im k_1\cdot x} H^{-2} \ex^{-|t_1-t_3|H}\ex^{-\im k_2\cdot x }H  (\vepsilon_2 \cdot x)\Gr\ra
 \no
  &  \times \la\Gr,  (\vepsilon_1 \cdot x) H  \ex^{\im k_1\cdot x} H^{-2}\ex^{-|t_2-t_4|H}\ex^{\im k_2\cdot x }H  (\vepsilon_2 \cdot x)\Gr\ra\ 
  \no
 &+\theta(t_1-t_2)\omega_1 \theta(t_3-t_4)\omega_2 \ex^{-\omega_1|t_1-t_2|} \ex^{-\omega_2|t_3-t_4|}
 \no
 &\times \Big(
 \la\Gr,  (\vepsilon_1 \cdot x) H  \ex^{-\im k_1\cdot x}H^{-2} \ex^{-|t_1-t_3|H}\ex^{-\im k_2\cdot x }H  (\vepsilon_2 \cdot x)\Gr\ra 
 \no
 &\times\delta(t_2-t_4)B_+(k_1, \lambda_1; k_2, \lambda_2)
 \no
 & + \la\Gr,  (\vepsilon_1 \cdot x) H  \ex^{\im k_1\cdot x} H^{-2}\ex^{-|t_2-t_4|H}\ex^{\im k_2\cdot x }H  (\vepsilon_2 \cdot x)\Gr\ra\ 
 \no
 & \times \delta(t_1-t_3)B_-(k_1, \lambda_1; k_2, \lambda_2)
 \Big)
 \no
 &+\ex^{-\omega_1|t_1-t_2|}\ex^{-\omega_2|t_3-t_4|}\delta(t_1-t_3)\delta(t_2-t_4) 
 \no
 &\times B_-(k_1, \lambda_1; k_2, \lambda_2)B_+(k_1, \lambda_1; k_2, \lambda_2) \Big\},
\end{align}
where
\begin{align}
&B_-(k_1, \lambda_1; k_2, \lambda_2)
\no
&=(\vepsilon_1\cdot \vepsilon_2)\la \Gr, \ex^{-\im(k_1+k_2)\cdot x}\Gr\ra-2\la \Gr, (\vepsilon_1\cdot x) H \ex^{-\im k_1\cdot x} H^{-1}\ex^{-\im k_2\cdot x}H (\vepsilon_2\cdot x)\Gr\ra,
\\
&B_+(k_1, \lambda_1; k_2, \lambda_2)
\no
&=(\vepsilon_1\cdot \vepsilon_2)\la \Gr, \ex^{\im(k_1+k_2)\cdot x}\Gr\ra-2\la \Gr, (\vepsilon_1\cdot x) H \ex^{\im k_1\cdot x} H^{-1}\ex^{\im k_2\cdot x}H (\vepsilon_2\cdot x)\Gr\ra
\end{align}

We use (\ref{Dec2ndCum}) and  collect the terms of $C_2(R, T)$ according to the number of their time-integrations, divide by $1/2T$, and take the limit as $T\to \infty$. To prepare
 for the limit $R\to \infty$,
we rescale the momentum integration as $k_j\leadsto k_j/R, j=1,2$, and introduce 
the unit vector $\hat{n}=r/R=(0,0,1)$. Note that $\vepsilon(k/R, \lambda)=\vepsilon(k, \lambda)$. 
 The two-time, three-time, and four-time integrations are treated separately.

\subsection{Two-time integrations}\label{TwoTime}
The sum of all terms involving  two-time integrations  is denoted by $I_2(R)$. One has 
\begin{align}
I_2(R)=&\int\dm k_1\dm k_2 |\hphi(k_1/R)|^2|\hphi(k_2/R)|^2 \ex^{\im (k_1+k_2)\cdot \n}
\no
& \times \Big[ \int \dm t\Big\{
R^{-2} (|k_1|^2 |k_2|^2)^{-1} \Big(
\big|
\la\Gr,  (1-\ex^{-\im k_1\cdot x/R})\ex^{-|t|H}(1-\ex^{-\im k_2\cdot x/R})\Gr\ra
\big|^2
\no
& -\big|
\la\Gr, (1-\ex^{-\im k_1\cdot x/R})\Gr\ra\la\Gr, (1-\ex^{-\im k\cdot x_2/R})\Gr\ra
\big|^2
\Big)
\no
&+2\sum_{\lambda_1} R^{-4} |k_2|^{-2} \la \Gr, (\vepsilon_1\cdot x) H \ex^{-\im k_1\cdot x/R}H^{-1}\ex^{-|t|H} (1-\ex^{-\im k_2\cdot x/R})\Gr\ra
\no
&\times \la \Gr, (\vepsilon_1\cdot x) H \ex^{\im k_1\cdot x/R} H^{-1}\ex^{-|t|H} (1-\ex^{\im k_2\cdot x/R})\Gr\ra
\no
&+\sum_{\lambda_1, \lambda_2}R^{-6} \la \Gr, (\vepsilon_1\cdot x) H\ex^{-\im k_1\cdot x/R}H^{-2} \ex^{-|t|H} \ex^{-\im k_2\cdot x/R} H(\vepsilon_2\cdot x)
\Gr\ra
\no
&\times \la \Gr, (\vepsilon_1\cdot x) H\ex^{\im k_1\cdot x/R}H^{-2} \ex^{-|t|H} \ex^{\im k_2\cdot x/R} H(\vepsilon_2\cdot x)
\Gr\ra
\Big\}
\no
&+\sum_{\lambda_1, \lambda_2}R^{-3} (2\omega_1\omega_2 (\omega_1+\omega_2))^{-1}
\no
&\times  B_-(k_1/R, \lambda_1; k_2/R, \lambda_2)B_+(k_1/R, \lambda_1; k_2/R, \lambda_2)
\Big]\label{TwoTime1}
\no
=& I_{2,1}(R)+I_{2,2}(R).
\end{align}

We consider the sum, $I_{2,1}(R)$,  of the first three terms and expand in $1/R$,  which yields  expectations of the form $\la \Gr, (a\cdot x) \ex^{-|t|H} (b\cdot x)\Gr\ra$ with $a,b\in \BbbR^3$. By rotation invariance
\begin{align}
\la \Gr, (a\cdot x) \ex^{-|t|H}(b\cdot x) \Gr\ra=(a\cdot b)\tfrac{1}{3}\la \Gr, x\cdot \ex^{-|t|H}x\Gr\ra.
\end{align} 
Using 
\begin{align}
\sum_{\lambda_1, \lambda_2}(\vepsilon_1\cdot \vepsilon_2)^2=1+(\hat{k}_1\cdot \hat{k}_2)^2,\ \ \sum_{\lambda_1} (\vepsilon_1\cdot \hat{k}_2)^2=1-(\hat{k}_1\cdot \hat{k}_2)^2,
\end{align}
 one arrives at the lowest order
\begin{align}
R^{-6}\big(\tfrac{1}{3}\la \Gr, x\cdot \ex^{-|t|H}x\Gr\ra\big)^2\big((\hat{k}_1\cdot \hat{k}_2)^2+2-2(\hat{k}_1\cdot \hat{k}_2)^2+1+(\hat{k}_1\cdot \hat{k}_2)^2\big). \label{Cal1}
\end{align}
The nonsmooth contributions, containing $(\hat{k}_1\cdot
\hat{k}_2)^2$, are canceled exactly, while  the smooth part, at this
order, is given by
\begin{align}
&I_{2,1}(R)
\no
&=3\int\dm t\int \dm k_1\dm k_2 |\hphi(k_1/R)|^2|\hphi(k_2/R)|^2\ex^{\im (k_1+k_2)\cdot \n} R^{-6} \big(\tfrac{1}{3}\la \Gr, x\cdot \ex^{-|t|H}x\Gr\ra\big)^2,
\end{align}
which inherits the rapid decay in $R$ from $\vphi$.  At the next order one picks up the quadratic contributions  $k_1^2, k_1\cdot k_2, k_2^2$ with coefficients integrable in $t$. By power counting  one arrives at 
$I_{2, 1}(R)\simeq R^{-8}$ as $R\to \infty$.

For the second summand, $I_{2,2}(R)$, we use that to  leading order in $1/R$, \begin{align}
B_{\pm}(k_1/R, \lambda_1; k_2/R, \lambda_2)= R^{-2} |k_1| |k_2| (\vepsilon_1\cdot\hat{ k}_2)   (\vepsilon_2\cdot \hat{k}_1)(\tfrac{1}{3}\la \Gr, x^2\Gr\ra)+\mathcal{O}(R^{-4}),
\end{align}
as in shown in Appendix \ref{GrEx}.
Using that $|\hphi(0)|^2=(2\pi)^{-3}$ one arrives at
\begin{align}
&I_{2,2}(R)
\no
&=  R^{-7}\big(\tfrac{1}{3} \la \Gr, x^2 \Gr\ra\big)^2(2\pi)^{-6}\frac{1}{2} \int \dm k_1 \dm k_2  (\omega_1\omega_2(\omega_1+\omega_2))^{-1}|k_1|^2 |k_2|^2
\no
&\times \ex^{\im (k_1+k_2)\cdot \n}
 (1-(\hat{k}_1\cdot
\hat{k}_2)^2)^2 
 +\mathcal{O}(R^{-9})
\no
&= \frac{128}{\pi}\big(\frac{1}{2\pi}\big)^2  e^{-4 } \alpha_{\mathrm{M}}^2 R^{-7}
+\mathcal{O}(R^{-9}).
\end{align}
The evaluation  of the numerical coefficient is discussed in Appendix \ref{NumC}.
  Altogether 
  \begin{align}
  \lim_{R\to \infty} R^7I_2(R)= \frac{128}{\pi}\big(\frac{1}{2\pi}\big)^2 e^{-4} \alpha_{\mathrm{M}}^2. \label{I2}
  \end{align}
\subsection{Three-time integrations}
The sum of all terms involving three-time integrations is denoted by $I_3(R)$. One obtains
\begin{align}
I_3(R)=&\int\dm t_1\dm t_2 \int \dm k_1\dm k_2 |\hphi(k_1/R)|^2|\hphi(k_2/R)|^2 \ex^{\im(k_1+k_2)\cdot \n}
\no
&\times \Big\{
-R^{-5} \omega_1 |k_2|^{-2} \ex^{-\omega_1|t_1-t_2|/R}
\no
&\times\Big( \sum_{\lambda_1}  \la \Gr, (\vepsilon_1\cdot x) H \ex^{-\im k_1\cdot x/R}H^{-1}
\ex^{-|t_1|H} (1-\ex^{-\im k_2\cdot x/R})\Gr\ra
\no
&\times  \la \Gr, (\vepsilon_1\cdot x) H \ex^{\im k_1\cdot x/R}H^{-1}
\ex^{-|t_2|H} (1-\ex^{\im k_2\cdot x/R})\Gr\ra\Big)
\no
&-R^{-7}\big(
\tfrac{1}{2}\omega_1 \ex^{-\omega_1|t_1-t_2|/R}+\tfrac{1}{2}\omega_2 \ex^{-\omega_2|t_1-t_2|/R}\big) \no
&\times \Big(\sum_{\lambda_1, \lambda_2}
\la \Gr, (\vepsilon_1\cdot x) H \ex^{-\im k_1\cdot x/R}H^{-2} \ex^{-|t_1|H}\ex^{-\im k_2\cdot x/R} H(\vepsilon_2\cdot x)\Gr\ra
\no
&\times 
\la \Gr, (\vepsilon_1\cdot x) H \ex^{\im k_1\cdot x/R}H^{-2} \ex^{-|t_2|H}\ex^{\im k_2\cdot x/R} H(\vepsilon_2\cdot x)\Gr\ra\Big)\nonumber
\end{align}

\begin{align}
&+R^{-6}\tfrac{1}{4} \theta(t_1) \ex^{-\omega_1 |t_1|/R} \theta(t_2)\ex^{-\omega_2 |t_2|/R}
\no
&\times\Big(\sum_{\lambda_1, \lambda_2} \big(
\la\Gr,  (\vepsilon_1 \cdot x) H  \ex^{-\im k_1\cdot x/R}H^{-2} \ex^{-|t_1-t_2|H}\ex^{-\im k_2\cdot x/R }H  (\vepsilon_2 \cdot x)\Gr\ra 
\no
&\times B_+(k_1, \lambda_1; k_2, \lambda_2)
\no
&+\la\Gr,  (\vepsilon_1 \cdot x) H  \ex^{\im k_1\cdot x/R}H^{-2} \ex^{-|t_1-t_2|H}\ex^{\im k_2\cdot x/R }H  (\vepsilon_2 \cdot x)\Gr\ra 
\no
&\times B_-(k_1, \lambda_1; k_2, \lambda_2)
 \big)\Big)
\Big\}\no
=& I_{3,1}(R)+I_{3,2}(R).
\end{align}

We expand the integrand in $1/R$ which yields
\begin{align}
&\int\dm t_1\dm t_2 \int \dm k_1\dm k_2 |\hphi(k_1/R)|^2|\hphi(k_2/R)|^2 \ex^{\im(k_1+k_2)\cdot \n}
\no
&\times \Big\{
-R^{-7} \omega_1  \ex^{-\omega_1|t_1-t_2|/R} (1-(\hat{k}_1\cdot\hat{k_2})^2)
\no
&\times \tfrac{1}{3}\la \Gr, x\cdot 
\ex^{-|t_1|H} x\Gr\ra \tfrac{1}{3}\la \Gr, x\cdot  
\ex^{-|t_2|H} x\Gr\ra
\no
&-R^{-7}\big(
\tfrac{1}{2}\omega_1 \ex^{-\omega_1|t_1-t_2|/R}+\tfrac{1}{2}\omega_2 \ex^{-\omega_2|t_1-t_2|/R}\big) (1+(\hat{k}_1\cdot \hat{k}_2)^2)
\no
&\times \tfrac{1}{3}\la \Gr, x\cdot  
\ex^{-|t_1|H} x\Gr\ra \tfrac{1}{3}\la \Gr, x\cdot  
\ex^{-|t_2|H} x\Gr\ra
\no
&+R^{-6} \tfrac{1}{4}\theta(t_1) \ex^{-\omega_1 |t_1|/R} \theta(t_2)\ex^{-\omega_2 |t_2|/R}
\no
&\times \Big(\sum_{\lambda_1, \lambda_2}
\tfrac{1}{3}\la \Gr, x\cdot  
\ex^{-|t_1-t_2|H} x\Gr\ra 
\no
&\times (\vepsilon_1\cdot \vepsilon_2)\big(
 B_+(k_1/R, \lambda_1; k_2/R, \lambda_2)
+ B_-(k_1/R, \lambda_1; k_2/R, \lambda_2)
 \big)\Big)
\Big\}\label{ThreeT1}
\no
=&-R^{-7}\int\dm t_1\dm t_2 \int\dm k_1\dm k_2 |\hphi(k_1/R)|^2|\hphi(k_2/R)|^2
\ex^{\im (k_1+k_2)\cdot \n} \omega_1 \ex^{-\omega_1|t_1-t_2|/R}
\no
&\times \tfrac{1}{3}\la \Gr, x\cdot 
\ex^{-|t_1|H} x\Gr\ra \tfrac{1}{3}\la \Gr, x\cdot  
\ex^{-|t_2|H} x\Gr\ra
\no
&+ R^{-6}\int\dm t_1\dm t_2 \int\dm k_1\dm k_2 |\hphi(k_1/R)|^2|\hphi(k_2/R)|^2\Big(
\sum_{\lambda_1, \lambda_2}
\no
&\times \ex^{\im (k_1+k_2)\cdot \n}\tfrac{1}{4}\theta(t_1)\ex^{-\omega_1|t_1|/R} \theta(t_2) \ex^{-\omega_2 |t_2|/R}
 \tfrac{1}{3}\la \Gr, x\cdot  
\ex^{-|t_1-t_2|H} x\Gr\ra 
\no
&\times (\vepsilon_1\cdot \vepsilon_2)\big(
 B_+(k_1/R, \lambda_1; k_2/R, \lambda_2)
+ B_-(k_1/R, \lambda_1; k_2/R,  \lambda_2)
 \big)\Big).
\end{align}
For the first summand the nonsmooth contributions are canceled
 exactly, as in Section \ref{TwoTime}, while the smooth contribtions decay rapidly
 since $\vphi$ does so. Its  next order  picks up an extra factor $R^{-2}$.  Thus
$I_{3,1}(R)\simeq R^{-9} $ for large $R$. For the second summand we perform explicitly the time integration with the result
\begin{align}
I_{3,2}(R)=&R^{-6} \int \dm k_1\dm k_2  |\hphi(k_1/R)|^2|\hphi(k_2/R)|^2\Big( 
\ex^{\im (k_1+k_2)\cdot \n} 
\no
&\times  R (\omega_1+\omega_2)^{-1}\tfrac{1}{3}
\la \Gr, x\cdot \big(H (H+R^{-1}\omega_1)^{-1} (H+R^{-1}\omega_2)^{-1}\big)
x\Gr\ra 
\no
&\times\sum_{\lambda_1, \lambda_2}
 (\vepsilon_1\cdot \vepsilon_2)\big(
 B_+(k_1/R, \lambda_1; k_2/R, \lambda_2)
+ B_-(k_1/R, \lambda_1; k_2/R, \lambda_2)\big)\Big).
\end{align}
Since $B_{\pm}=R^{-2} |k_1| |k_2|(\vepsilon_1\cdot \hat{k}_2) (\vepsilon_2\cdot \hat{k}_1)(\tfrac{1}{3}\la \Gr, x^2\Gr\ra)+\mathcal{O}(R^{-4})$, as shown in  Appendix \ref{GrEx}, we conclude that 
\begin{align}
I_{3,2}(R)=&-R^{-7}\big(\tfrac{1}{3}\la \Gr, x^2\Gr\big\ra\big)\big(\tfrac{1}{3}\la \Gr, x\cdot H^{-1}x\Gr\ra\big)
\no
&\times 2 (2\pi)^{-6} \int \dm k_1 \dm  k_2\,  \ex^{\im (k_1+k_2)\cdot \hat{n}} (\omega_1\omega_2(\omega_1+\omega_2))^{-1}
\no
&\times |k_1|^2 |k_2|^2 (\hat{k}_1\cdot \hat{k}_2)
(1-(\hat{k}_1\cdot \hat{k}_2)^2)  
  +\mathcal{O}(R^{-9})
\no
=&\frac{52}{\pi}\big(\frac{1}{2\pi}\big)^2 e^{-4}\alpha_{\mathrm{E}}\alpha_{\mathrm{M}} R^{-7}+\mathcal{O}(R^{-9}).
\end{align}
  The evaluation of the numerical coefficient is discussed  in Appendix \ref{NumC}. Altogether one has 
  \begin{align}
  \lim_{R\to \infty} R^7 I_3(R)=\frac{52}{\pi}\big(\frac{1}{2\pi}\big)^2  e^{-4 } \alpha_{\mathrm{E}}\alpha_{\mathrm{M}}. \label{I3Lim}
    \end{align}

\subsection{Four-time integrations}
There is only a single term with four-time integrations, namely
\begin{align}
I_4(R)=&  \int\dm t_1\dm t_2\dm t_3 \sum_{\lambda_1, \lambda_2}\int\dm k_1 \dm k_2 |\hphi_1|^2|\hphi_2|^2
 \ex^{\im (k_1+k_2)\cdot r} 
 \no
 &\times \frac{1}{4}\omega_1\omega_2 \ex^{-\omega_1|t_1-t_2+t_3|}\ex^{-\omega_2 |t_3|}
 \no
 &\times \la \Gr, (\vepsilon_1\cdot x) H \ex^{-\im k_1\cdot x}H^{-2}\ex^{-|t_1|H} \ex^{-\im k_2\cdot x} H (\vepsilon_2\cdot x)\Gr\ra
 \no
 &\times \la \Gr, (\vepsilon_1\cdot x) H \ex^{\im k_1\cdot x}H^{-2}\ex^{-|t_2|H} \ex^{\im k_2\cdot x} H (\vepsilon_2\cdot x)\Gr\ra.
 \end{align}
We rescale $k_j \leadsto k_j/R$ as before and in addition $t_3\leadsto t_3R$. Then the exponentially decaying  terms are 
\begin{align}
\ex^{-\omega_1|t_3+(t_1-t_2)/R|}\ex^{-\omega_2 |t_3|}\simeq \ex^{-\omega_1 |t_3|} \ex^{-\omega_2 |t_3|}
\end{align}
for large $R$. Expanding in $1/R$ yields
\begin{align}
I_4(R)=  &R^{-7}  \big(\tfrac{1}{3} \la \Gr, x\cdot H^{-1}x\Gr\ra\big)^2 2 \int_0^{\infty}\dm t \int \dm k_1 \dm k_2 |\hphi(k_1/R)|^2|\hphi(k_2/R)|^2
 \no
 &\times \ex^{\im (k_1+k_2)\cdot \n} 
  \omega_1\omega_2 \ex^{-t (\omega_1+\omega_2)} (1+(\hat{k}_1\cdot \hat{k}_2)^2)  +\mathcal{O}(R^{-9})
  \no
= &R^{-7}  (2\pi)^{-6}\big(\tfrac{1}{3} \la \Gr, x\cdot H^{-1}x\Gr\ra\big)^2  2\int \dm k_1 \dm k_2 
   \no
 &\times \ex^{\im (k_1+k_2)\cdot \n} \frac{\omega_1\omega_2}{\omega_1+\omega_2} (1+(\hat{k}_1\cdot \hat{k}_2)^2)+\mathcal{O}(R^{-9})
 \no
 =& R^{-7} \big(\frac{1}{2\pi}\big)^2 e^{-4}\frac{23}{8\pi} \alpha_{\mathrm{E}}^2+\mathcal{O}(R^{-9}). \label{I4}
  \end{align}
  The prefactor of $R^{-7}$ agrees with  the one computed already by
  Casimir and Polder.
  
  \subsection{Sum of all terms}
    We add the limits listed in (\ref{I2}), (\ref{I3Lim}),  and (\ref{I4}), which yields
  \begin{align}
 \kappa(e)= \lim_{R\to \infty} R^7 \Big(\lim_{T\to \infty} \frac{1}{2T}e^4 \frac{1}{2}C_2(R,T)\Big)=\big(\frac{1}{2\pi}\big)^2 \Big(\frac{23}{16\pi}\alpha_{\mathrm{E}}^2
  +\frac{26}{\pi}\alpha_{\mathrm{E}}\alpha_{\mathrm{M}}
  +\frac{64}{\pi}\alpha_{\mathrm{M}}^2\Big).\label{SumAll}
    \end{align}

   \subsection{Comparison with previous results}

Magnetic contributions  to the  $-R^{-7}$ decay were first considered by Feinberg and Sucher \cite{FS0, FS} and by Boyer \cite{Boyer} .
They find that $\alpha^2_{\mathrm{E,at }}$ and $\alpha_{\mathrm{M, at}}^2$ have the same coefficient, namely  $23/16\pi$, while  the one of $\alpha_{\mathrm{E, at}}
\alpha_{\mathrm{M, at}}$ is $7/8\pi$. This raises  the issue on  the origin for the discrepancy.

The $\alpha_{\mathrm{E, at}}^2$ term  can be most easily obtained through the dipole approximated hamiltonian
\begin{align}
H_{\mathrm{dip}}=&\frac{1}{2} (p_1-eA(0))^2-e^2 V(x_1)+\frac{1}{2} (p_2-e A(r))^2-e^2 V(x_2)+\Hf
\no &+ e^2 V_R(x_1, x_2). \label{Dipole}
\end{align}    
   One shifts $p_1$ by $eA(0)$ and  $p_2$ by $eA(r)$ through the unitary
   $U=\exp[-\im (x_1 \cdot A(0) +x_2 \cdot A(r))]$.  Then
  \begin{align}
  U H_{\mathrm{dip}} U^{-1} =& \frac{1}{2}p_1^2-e^2 V(x_1)+\frac{1}{2}p_2^2-e^2 V(x_2)+\Hf
  \no
  &-e x_1\cdot E_{\perp}(0)-e x_2\cdot E_{\perp}(r)+e^2 V_R(x_1, x_2)
  \no
  &+e^2 \int \mathrm{d}k\, |\hat{\vphi}(k)|^2 \Big(\frac{1}{3} x_1^2+\frac{1}{3}x_2^2+\ex^{\im k\cdot r}\big(
  x_1\cdot x_2-|k|^{-2} (x_1\cdot k)(x_2\cdot k)
  \big)\Big),\label{UniTra}
  \end{align}
   where $E_{\perp}(r)$ is the quantized transverse electric field.
   Note that the long range part of $V_R$ is cancelled. From the $4$-th order  perturbation in $-e x_1\cdot E_{\perp}(0)-ex_2 \cdot E_{\perp}(r)$
    one obtains the Casimir-Polder result. 
    
    To include magnetic effects one proceeds to the next order of  the multipole expansion and defines
    \begin{align}
    H_{\mathrm{mul}}=&\frac{1}{2} \big(p_1-eA(0)-e x_1\cdot \nabla_r A(0)\big)^2-e^2 V(x_1)
    \no
    &+\frac{1}{2} \big(p_2-e A(r)-e x_2\cdot \nabla_r A(r)  \big)^2-e^2 V(x_2)
\no &+\Hf+ e^2 V_R(x_1, x_2),
  \end{align}
    which,
   as before, is unitarily transformed  to 
   \begin{align}
  U H_{\mathrm{mul}}U^{-1}=&\frac{1}{2} \big(p_1-e x_1\cdot \nabla_r A(0)\big)^2-e^2 V(x_1)
    \no
    &+\frac{1}{2} \big(p_2-e x_2\cdot \nabla_r A(r)  \big)^2-e^2 V(x_2)
\no &+\Hf
-e x_1\cdot E_{\perp}(0)-e x_2\cdot E_{\perp}(r)+e^2 V_R^{\mathrm{cor}}(x_1, x_2), \label{UniTra2}
   \end{align}
   where $V_R^{\mathrm{cor}}$ is the interaction potential  from (\ref{UniTra}). Note that $\nabla_x\wedge (x\cdot \nabla_r A(r))=B(r)=\nabla_r\wedge A(r)$.
   Thus  the hamiltonian (\ref{UniTra2}) clearly displays the fluctuating  electric  and magnetic  fields. Expanding in the terms proportional to 
   $\nabla_r A(0), \\
   \nabla_r A(r), E_{\perp}(0), E_{\perp}(r)$ to $4$-th order  leads to an energy with a large $R$ asymptotics  in agreement  with (\ref{SumAll}).
   In  spirit  the authors of \cite{FS0, FS, Boyer} make  a further gauge transformation through  the unitary
   $
   \exp[ -\im (x_1\cdot \nabla_r (x_1\cdot A(0))+x_2 \cdot \nabla_r(x_2\cdot A(r)))].
   $   
   Thereby $x_1\cdot \nabla_rA(0)$ is transformed  to $\frac{1}{2} x_1\wedge B(0)$ and correspondingly  for  $x_2\cdot \nabla_r A(0)$.
   In addition there are terms coming from the shifting  of $E_{\perp}$. In the $4$-th order perturbation only  the former terms  are  taken into account. More
   precisely 
    the term
   $\la \Gr_{\mathrm{at}}\otimes \Omega, (x_1\wedge B(0))^2 P H^{-1}  P (x_2\wedge B(r))^2\Gr_{\mathrm{at}}\otimes \Omega\ra$, which 
   yields indeed the integrand (\ref{SS1}) and thus $23/16\pi$ for  the prefactor. Here $P=\one-|\Gr \ra\la \Gr|$ and $\Omega$  is the  Fock vacuum.
    For the cross-term they use 
   $
   \la \Gr_{\mathrm{at}}\otimes \Omega, x_1\cdot E_{\perp}(0) P H^{-1} P  x_1\cdot E_{\perp}(0) P H^{-1} P  (x_2\wedge B(r))^2\Gr_{\mathrm{at}}\otimes \Omega\ra$,
    which  yields the integral (\ref{SS2}) with the term $(1-(\hat{k}_1\cdot \hat{k}_2)^2)$ omitted   and  thus $7/8\pi$ for the prefactor of $\alpha_{\mathrm{E, at}} \alpha_{\mathrm{M, at}}$.

   \section{The  dimensionless strength}\setcounter{equation}{0}
\label{DL}
   
   We restore the physical units in $H$ of (\ref{1PL}). Then 
   \begin{align}
   H=\frac{1}{2m} \big(p-\tfrac{e}{c} A(x)\big)^2-e^2 V(x)+\Hf
   \end{align}
   with $p=-\im \hbar \nabla_x$,
   \begin{align}
   \Hf=\sum_{\lambda=1,2} \int_{\BbbR^3} \dm k\, \hbar c  |k| a(k, \lambda)^* a(k, \lambda),
   \end{align}
   and 
   \begin{align}
    A(x)=\sum_{\lambda=1,2} \int_{\BbbR^3} \dm k \sqrt{c/2|k|} \hat{\vphi}(k)\vepsilon(k, \lambda) \big(\ex^{\im k\cdot x}a(k, \lambda)+\ex^{-\im k\cdot x}a(k, \lambda)^*\big) . 
    \end{align}
   $H$ is   transformed to atomic units through the canonical transformation $U$ defined as 
   \begin{align}
   U^* a(k, \lambda)U&=(\alpha^{-2} \lambda_{\mathrm{c}})^{3/2} a(\alpha^{-2} \lambda_{\mathrm{c}} k, \lambda),
   \no
   U^* x U& =\alpha^{-1}r_{\mathrm{B}}x, \ \ \ U^* pU =\alpha r _{\mathrm{B}}^{-1} p.
   \end{align}
   Then 
   \begin{align}
   U^* H U&= \alpha^2 m c^2 \Big(
   \frac{1}{2} \big(
   -\im \nabla_x - \sqrt{4\pi} \alpha^{3/2} \tilde{A}(\alpha x)
   \big)^2   -  \tilde{V}(x) +\tilde{H}_{\mathrm{f}}\Big)
   \no
   &= \alpha^2 m c^2 \tilde{H},
   \end{align}
   where
   \begin{align}
   \tilde{H}_{\mathrm{f}}= \sum_{\lambda=1,2} \int_{\BbbR^3} \dm k\,   |k| a(k, \lambda)^* a(k, \lambda) \label{THf}
   \end{align}
    and 
    \begin{align}
    \tilde{V}(x)&=4\pi \int_{\BbbR^3}\dm k\,   \hat{\vphi}(\alpha^2 \lambda_{\mathrm{c}}^{-1} k)^2 |k|^{-2} \ex^{-\im k\cdot x},\label{TV}
    \\
         \tilde{A}(x)&=\sum_{\lambda=1,2}\int_{\BbbR^3} \dm k\,    \frac{\hat{\vphi}(\alpha^2 \lambda_{\mathrm{c}}^{-1} k)     }{\sqrt{2|k|}}\vepsilon(k, \lambda) \big(\ex^{\im k\cdot x}a(k, \lambda)+\ex^{-\im k\cdot x}a(k, \lambda)^*\big) .    \label{TA}
          \end{align}
   Energy is now in units of $\alpha^2 m c^2$  and distances are in units of the Bohr radius $r_{\mathrm{B}}$.
   
   From (\ref{THf}), (\ref{TA}) the photon propagator in atomic units is obtained as 
   
   \begin{align}
\tilde{W}_R(q,t)=4\pi 
\int_{\BbbR^3}\mathrm{d}k\,
|\hat{\vphi}(\alpha^2 \lambda_{\mathrm{c}}^{-1}k)|^2\frac{1}{2|k|}
(\one-| \hat{k}\ra\la\hat{k}|)\, \ex^{-\alpha^{-1} |k||t|}\ex^{\im k\cdot
  r}\ex^{-\im k\cdot q}.
   \end{align}
   Comparing with (\ref{Propagator}) this amounts to replacing  $e^2$ by $4\pi$ and $t$ by $t/\alpha$. One merely has  to follow through
    this change in the computation of Section \ref{Computation}. The final result is 
    \begin{align}
    E(R)-2E=-\kappa (\alpha, \lambda_{\mathrm{c}} \Lambda) \alpha m c^2 (R/r_{\mathrm{B}})^{-7}
    \end{align} 
    for large $R$ with  the strength $\kappa$ obtained as 
    \begin{align}
    \kappa (\alpha, \lambda_{\mathrm{c}}\Lambda)= \frac{4}{\pi} \Big(\frac{23}{16}\tilde{\alpha}_{\mathrm{E}}^2
  +26\tilde{\alpha}_{\mathrm{E}}\tilde{\alpha}_{\mathrm{M}}
  +64\tilde{\alpha}_{\mathrm{M}}^2\Big)    .
    \end{align}
    Here,  the dimensionless electric and magnetic dipole  moments are  
    \begin{align}
    \tilde{\alpha}_{\mathrm{E}}=2 \big(\tfrac{1}{3} \la \Gr, x\cdot (\tilde{H}-E)^{-1} x\Gr\ra\big)
    \end{align}
    and 
    \begin{align}
    \tilde{\alpha}_{\mathrm{M}} =-\alpha^2 \frac{1}{4} \big(\tfrac{1}{3} \la \Gr, x^2\Gr\ra\big)
    \end{align}
    with $\Gr$  the ground state of $\tilde{H}$, $\tilde{H}\Gr=E\Gr$. The ultraviolet cut-off is implemented by  replacing $\hat{\vphi}(k)$ by
     the form factor $\hat{\vphi}_1(k/ \Lambda )$: $\hat{\vphi}_1$ decreases rapidly at $|k|=1$ from $(2\pi)^{-3/2}$ to $0$.

    Following Bethe \cite{Bethe} a physically reasonable choice  is $\lambda_{\mathrm{c}} \Lambda=1$.  One would like to remove the ultraviolet cut-off 
    through $\Lambda\to \infty$.
    But  this limit  is not well understood. In any case, the bare mass $m$ would  have to  substituted by the renormalized mass.  For fixed $\Lambda,\ 
      \lambda_{\mathrm{c}}\Lambda =1$, in the limit $\alpha\to 0$,
    $\tilde{H}$ decouples from the radiation field and $\kappa(\alpha,
    \lambda_{\mathrm{c}} \Lambda)$ tends to the strength obtained by  Casimir  and Polder. Then $\tilde{\alpha}_{\mathrm{E}}\simeq 9/4$ and $\tilde{\alpha}_{\mathrm{M}}\simeq -\alpha^2$ with  quality for the strict Coulomb potential.
      For a systematic  expansion in $\alpha$ one would need the correction of
    order $\alpha$ to $\tilde{\alpha}_{\mathrm{E}}$ and to  the bare mass $m$.

\section{Conclusions and Outlook}\setcounter{equation}{0}
The reader may wonder  how much  is still missing to a complete proof.
For  the full model, a central difficulty is that the ground state process $q(t)$ is not so well under control. Since $H$ has no spectral gap, the two-point correlation has a slow decay,
presumably as   $\E[ q(t)\cdot q(0)] \simeq |t|^{-4}$ for large $t$. This means that  higher order cumulants are difficult 
to control. In  fact, we cannot even prove  that $\alpha_{\mathrm{E}}<\infty$. 

 From the statistical mechanics point of view an interesting case  would be to replace the ground state process $q(t)$  by $q_{\mathrm{at}}(t)$, namely
  the one  governed by $H_{\mathrm{at}}$. This process is
  exponentially  mixing which should  help  in controlling  the error
  term.
  On the other hand, the double stochastic integral in the action
  causes extra difficulties.
  Unfortunately there is no obvious hamiltonian corresponding to this approximation.
  
  A further variant   would be  the dipole approximation of equation (\ref{Dipole}).
   In the path integral   (\ref{PathBindingEr})  this corresponds to replacing $W_R(x, t)$ by $W(0, t)$.
    The effective action is quadratic in $\dot{q}_j(t)$ and the partial time integration can be easily implemented.
    It results in a diagonal  term which  cancels the slowly decaying  part of $V_R$ and the remainder  action is given by
   $\int \dm s \dm t q(s)\cdot \ddot{W}_R(0, s-t) q(t)$. Since $\ddot{W}_R(0, t)\simeq t^{-4}$ 
     for large $t$, the difficulties
mentioned above remain. To have an exactly solvable  Gaussian  model,
on top one would have to use the  quadratic approximation for
$V$ and $V_R$. 
 
For ground state properties a powerful method is the Feshbach
projection together with  a successive integration over high $k$-modes
of the  radiation 
field \cite{BCFS, GH}. It would be interesting to understand whether this
technique could be  used  for  a  rigorous  control on the van der
Waals potential. 

\begin{flushleft}
{\bf Acknowledgement:} We are grateful to P. Milonni for very useful
hints on the literature.
\end{flushleft}

\appendix
\section{Proof of (i) to (v)}\setcounter{equation}{0}
\label{(i)to(v)}
The proof of (i) to  (v) is based on the identity
\begin{align}
\E\Big[\prod_{j=1}^m f_j(q(t_j))\Big]=\la \Gr, f_1(x) \ex^{-|t_1-t_2|(H-E)}\cdots f_{m-1}(x) \ex^{-|t_{m-1}-t_m|(H-E)}f_m(x)\Gr\ra
\end{align}
for the time order $t_1<t_2<\cdots< t_m$. $\dot{q}(t)\dm t$ is the Ito stochastic integral as defined  through the forward discretization. For this purpose
 we introduce the lattice spacing $\delta$ and let $[t]_{\delta}$ be the integer part of $t$ modulo $\delta$. Then $\dot{q}(t)= \lim_{\delta\to 0}\delta^{-1}(q([t]_{\delta}+\delta)-q([t]_{\delta}))$. We only establish (ii) and (v). The other items are proved by the same procedure.\\
 ad (ii): Using stationarity,
 \\
 \begin{align}
 \E[
 \ex^{\im k\cdot q(t)}\vepsilon\cdot \dot{q}(t)
 ]&=\lim_{\delta\to 0} \frac{1}{\delta}\E[\ex^{\im k\cdot q(0)} \vepsilon\cdot (q(\delta)-q(0))]
 \no
 &=\lim_{\delta\to 0}\frac{1}{\delta} \la \Gr, \ex^{\im k\cdot x}(\ex^{-\delta (H-E)}-1) (\vepsilon\cdot x)\Gr\ra
 \no
 &=-\la \Gr, \ex^{\im k\cdot x} (H-E)(\vepsilon\cdot x)\Gr\ra,
 \end{align} 
which vanishes as proved in  
  Proposition \ref{VGREXP}. 
 \medskip\\
ad (v): For $t\neq s$ we proceed as in ad (ii), which  yields the first summand of (v). For the ``diagonal  part" one has, for small $\delta$, 
\begin{align}
&\E\big[
(\vepsilon_1\cdot (q(\delta)-q(0))) \ex^{-\im k_1\cdot q(0)} \ex^{-\im k_2\cdot q(0)} (\vepsilon_2\cdot (q(\delta)-q(0)))
\big]
\no
&= \la \Gr, \ex^{-\im k_1\cdot x}\ex^{-\im k_2\cdot x}\big(\ex^{-\delta (H-E)} (\vepsilon_1\cdot x)(\vepsilon_2\cdot x)-(\vepsilon_1\cdot x)\ex^{-\delta (H-E)}
(\vepsilon_2\cdot x)
\no
&- (\vepsilon_1\cdot x) \ex^{-\delta (H-E)}(\vepsilon_2\cdot x)+(\vepsilon_1\cdot x)(\vepsilon_2\cdot x)\ex^{-\delta (H-E)}
\big)\Gr\ra
\no
&= -\delta \la \Gr, \ex^{-\im k_1\cdot x} \ex^{-\im k_2\cdot x} [[\delta^{-1}(1-\ex^{-\delta(H-E)}), (\vepsilon_1\cdot x)], (\vepsilon_2\cdot x)]\Gr\ra
\no
&\simeq \delta (\vepsilon_1\cdot \vepsilon_2)\la \Gr, \ex^{-\im k_1\cdot x}\ex^{-\im k_2\cdot x}\Gr\ra.
\end{align}

\section{Ground state expectations}\setcounter{equation}{0}
\label{GrEx}
In this appendix, we mainly work in the Schr\"odinger representation $L^2(\BbbR^3_x)\otimes \mathfrak{F}=L^2(\BbbR^3_x)\otimes L^2(Q)$, see  \cite{Spohn, Hiroshima1}. In this representation it holds
\begin{itemize}
\item[{\rm (a)}] $A(x)$ is a real-valued multiplication operator.
\item[{\rm (b)}] Let $\vartheta=\Gamma(\ex^{\im \pi/2})$. Then $\vartheta \ex^{-tH}\vartheta^{-1}$ is positivity improving. Hence
$\TGr=\vartheta \Gr$ is strictly positive.
\end{itemize}
Here for a unitary operator $U$, $\Gamma(U)$ is defined by
\[
\Gamma(U)a(f_1)^*\cdots a(f_n)^*\Omega=a(Uf_1)^*\cdots a(Uf_n)^*\Omega
\]
for all $n\in \BbbN$ and $f_1, \dots, f_n\in \BbbC^2\otimes L^2(\BbbR^3)$, where $\Omega$ is the Fock vacuum.    A linear operator $A$ ``improves the positivity" if, for all  $0\le f\in (L^2(\BbbR^3_x)\otimes L^2(Q))\backslash \{0\}$, $\ex^{-t A}f>0$ for all $t>0$. 
Let $J$ be the natural conjugation in $L^2(\BbbR^3_x)\otimes L^2(Q)$, namely, $J\Psi=\overline{\Psi}$ for all $\Psi\in L^2(\BbbR^3_x)\otimes L^2(Q)$.
A linear operator $A$ is called  to be $J$-real if $A$ commutes with $J$, i.e.,  $JA=AJ$.
Then, from (b), it follows that \\

(c) $\tilde{H}=\vartheta H\vartheta^{-1}$ is a $J$-real operator, i.e., $J\tilde{H}=\tilde{H}J$.\\

\begin{lemm}\label{VanisngGrlemm}(Vanishing ground state expectation I).
Let $F(x)$ a measurable function on $\BbbR^3_x$ such that  $\Gr\in \D(F)$. Then one has 
\begin{align}
\la \Gr, F(x) A(x) \Gr\ra=0. \label{VGr1}
\end{align}
\end{lemm}
\begin{rem}{\rm
From the proof one infers the stronger property
\[
\la \Gr, F(x) A(x)_{i_1}\ex^{-s_1 H}\cdots \ex^{-s_{2n}H} A(x)_{i_{2n+1}} \Gr \ra=0
\]
for all $n\in \{0\}\cup \BbbN, s_1, \dots, s_{2n}>0$ and $F\in L^{\infty}(\BbbR^3)$. (Note that we have to use (c).)
\rm}
\end{rem}
{\it Proof.} Note that 
\begin{align}
J \vartheta =\Gamma(\ex^{\im \pi}) \vartheta J.\label{JCommu}
\end{align}
Since  $A(x)$ is a real-valued multiplication operator,  one has  $A(x)J=JA(x)$. Hence using  (\ref{JCommu}), one sees
that $E(x)=\vartheta A(x)\vartheta^{-1}$ is  purely $J$-imaginary, that is, $JE(x)=-E(x)J$.
Remark that 
\[
\la \Gr, A(x) \Gr\ra=\la \TGr, E(x) \TGr\ra. 
\]
Since $\TGr$ is real-valued  by (b), the right hand side is purely imaginary. On the other hand, by the self-adjointness of $A(x)$, the left hand side
is real. Thus the only possibility is $\la \Gr, A(x) \Gr\ra=0$. Similarly one can show that 
\[
\la \Gr, \Re F (x)A(x) \Gr\ra=0=\la \Gr, \Im F(x)A(x) \Gr\ra
\]
which implies the assertion.  $\Box$

 \begin{lemm}\label{VanisngGrlemm2}(Vanishing ground state expectation II). 
 One has the following:
 \begin{itemize}
 \item[{\rm i)}] For all $a\in \BbbR^3$,
 $
 \la \Gr, (a\cdot x) ^{2n+1}\Gr\ra=0.
 $
  \item[{\rm ii)}] If $m+n$ is odd, then
  $
  \la \Gr, (a\cdot x) H (b\cdot x)^m H^{-1} (c\cdot x)^n H(d\cdot x)\Gr\ra=0
  $
  for all $a,b,c, d\in \BbbR^3$.
 \end{itemize}
 \end{lemm}
{\it Proof.}
For the  proof, we return  to the Fock representation $L^2(\BbbR^3_x)\otimes \Fock$. 
Let $J_2$ be the involution defined by 
\begin{align}
J_2  \Psi=\Sumoplus \overline{\Psi}_n(-x; k_1,\lambda_1,\dots, k_n, \lambda_n )
\end{align}
for all $\Psi=\sum_{n\ge 0}^{\oplus}\Psi_n(x; k_1, \lambda_1,\dots, k_n, \lambda_n)\in L^2(\BbbR^3_x)\otimes \Fock$.
  Then as proved in \cite{LMS}, we can check that 
  \begin{align}
  J_2 x&=-xJ_2,
  \\
  J_2 H&=HJ_2, \label{J2H}
  \end{align}
namely, $x$ is purely $J_2$-imaginary and $H$ is $J_2$-real.  Note  that $J_2\Gr=\ex^{\im \theta}\Gr$ for some $\theta\in [0, 2\pi)$ by (\ref{J2H}) and the uniqueness of the ground state.  
Hence 
\begin{align}
\la \Gr, (a\cdot x) ^{2n+1}\Gr\ra&=-\la J_2  \Gr, (a\cdot x) ^{2n+1}J_2 \Gr\ra
\no
&=-\la \Gr, (a\cdot x) ^{2n+1}\Gr\ra,
\end{align}
which implies $\la \Gr, (a\cdot x) ^{2n+1}\Gr\ra=0$. Similarly if $m+n$ is odd, then
\begin{align}
&\la \Gr, (a\cdot x) H (b\cdot x)^m H^{-1} (c\cdot x)^n H(d\cdot x)\Gr\ra
\no
&=-\la J_2\Gr, (d\cdot x) H (c\cdot x)^m H^{-1} (b\cdot x)^n H(a\cdot x)J_2 \Gr\ra\no
&=-\la \Gr, (d\cdot x) H (c\cdot x)^n H^{-1} (b\cdot x)^m H(a\cdot x) \Gr\ra.\label{VGr2Eq}
\end{align}
On the other hand, in the Schr\"odinger representation, we can see that 
\begin{align}
&\la \tilde{\Gr}, (d\cdot x) \tilde{H} (c\cdot x)^n \tilde{H}^{-1} (b\cdot x)^m \tilde{H}(a\cdot x) \tilde{\Gr}\ra
\no
&=\la \tilde{\Gr}, (a\cdot x) \tilde{H} (b\cdot x)^m \tilde{H}^{-1} (c\cdot x)^n \tilde{H}(d\cdot x)\tilde{\Gr}\ra
\end{align}
 because every operator appearing  in the expectation is  $J$-real. Therefore
 \begin{align}
&\la \Gr, (d\cdot x) H (c\cdot x)^n H^{-1} (b\cdot x)^m H(a\cdot x) \Gr\ra
\no
&=\la \Gr, (a\cdot x) H (b\cdot x)^m H^{-1} (c\cdot x)^n H(d\cdot x)\Gr\ra.
\end{align}
 Combining this with (\ref{VGr2Eq}), we  conclude ii). $\Box$

\begin{Prop}\label{VGREXP}
One has the following
\begin{itemize}
\item[{\rm i)}] $\displaystyle  \la \Gr, \ex^{\im k\cdot x} H(\vepsilon(k, \lambda)\cdot x)\Gr\ra=0$,
\item[{\rm ii)}] $\displaystyle B_{\pm}(k_1/R, \lambda_1; k_2/R, \lambda_2)=R^{-2} |k_1| |k_2|(\vepsilon_1\cdot \hat{k}_2)(\vepsilon_2\cdot \hat{k}_1)(\tfrac{1}{3}\la \Gr, x^2\Gr\ra)+\mathcal{O}(R^{-4})$ as $R\to \infty$.
\end{itemize}
\end{Prop}
{\it Proof.} ad  i)  By (\ref{Identity3}) below  and by  $\la\Gr, \ex^{-\im k\cdot x} A(x)\Gr\ra=0$ according to  Lemma \ref{VanisngGrlemm}, one has 
\begin{align*}
\la \Gr, \ex^{\im k\cdot x} H(\vepsilon(k, \lambda)\cdot x)\Gr\ra=&-\im \la \Gr, \ex^{\im k\cdot x}(\vepsilon(k, \lambda)\cdot (p-e A(x)))\Gr\ra\\
=&-\im \la \TGr, \ex^{\im k\cdot x}(\vepsilon(k, \lambda)\cdot p)\TGr\ra\\
=&- \frac{\im}{2}\int \dm x\dm \mu (A) \ex^{\im k\cdot x} \vepsilon(k, \lambda)\cdot (-\im \nabla_x\TGr^2)(x, A)\\
=&-  \frac{\im}{2} \int \dm x\dm \mu (A) \ex^{\im k\cdot x} (\vepsilon(k, \lambda)\cdot  k ) \TGr^2(x, A)\\
=&0,
\end{align*}
since  $k\cdot \vepsilon(k, \lambda)=0$.

ad ii)
By Lemma \ref{VanisngGrlemm2},  the order $R^{-1}$ vanishes and 
it suffices to check the order $R^0$ and $R^{-2}$. \\

ii-a) order $R^0$.   One has 
\begin{align}
\lim_{R\to \infty}B_-(k_1/R, \lambda_1; k_2/R, \lambda_2)=(\vepsilon_1\cdot \vepsilon_2)-2 \la \Gr, (\vepsilon_1\cdot x) H (\vepsilon_2\cdot x)\Gr\ra.
\end{align}
Furthermore, using the identities 
 \begin{align}
[a\cdot \tilde{p}, b\cdot x]&=-\im a\cdot b,\\
[H, a\cdot x]&=-\im a\cdot \tilde{p},\\
 H (a\cdot x)\Gr&=-\im a\cdot \tilde{p}\Gr\label{Identity3}
\end{align}
for all $a,b\in \BbbC^3$, where $\tilde{p}=p-eA(x)$, one obtains
\begin{align} 
2 \la \Gr, (\vepsilon_1\cdot x) H (\vepsilon_2\cdot x)\Gr\ra=\im \la \Gr, \big(
(\vepsilon_1\cdot \tilde{p}) (\vepsilon_2\cdot x)- (\vepsilon_1\cdot x) (\vepsilon_2\cdot \tilde{p})
\big)\Gr\ra
=(\vepsilon_1\cdot \vepsilon_2).
\end{align}
Remark that we have used (\ref{VGr1}) to conclude that $\la \Gr, (\vepsilon_1\cdot A(x)) (\vepsilon_2\cdot x)\Gr\ra=0\\=\la \Gr, (\vepsilon_1\cdot x)(\vepsilon_2\cdot A(x)) \Gr\ra$.
  \medskip\\
  ii-b) order $R^{-2}$. We expand as 
  \begin{align}
  R^2 B_-(k_1/R, \lambda_1;  k_2/R, \lambda_2)=c_1 k_1^2+c_2k_2^2 +d (k_1, \lambda_1 ; k_2, \lambda_2)+\mathcal{O}(R^{-2}).
  \end{align}
  For $c_1k_1^2$, the expansion coefficient is 
  \begin{align}
  (\vepsilon_1\cdot \vepsilon_2)\la \Gr, (k_1\cdot x)^2 \Gr\ra-2 \la \Gr, (\vepsilon_1\cdot x)H (k_1\cdot x)^2 (\vepsilon_2\cdot x)\Gr\ra.
  \end{align}
  Using  (\ref{VGr1}) and that $\TGr$ and $\tilde{H}$ can be chosen as real, for the second term one obtains 
  \begin{align}
  &\la \Gr, (\vepsilon_1\cdot x) H (k_1\cdot x)^2 (\vepsilon_2\cdot x)\Gr\ra +\la \Gr, (\vepsilon_2\cdot x) (k_1\cdot x)^2 H (\vepsilon_1\cdot x)\Gr\ra
  \no
  &= \im \la \Gr, (\vepsilon_1\cdot \tilde{p}) (k_1\cdot x)^2 (\vepsilon_2\cdot x)\Gr\ra-\im \la \Gr, (\vepsilon_2\cdot x) (k_1\cdot x)^2 (\vepsilon_1\cdot \tilde{p})\Gr\ra
  \no
  &=\im \la \Gr, (k_1\cdot x)^2 [(\vepsilon_1\cdot\tilde{p}), (\vepsilon_2\cdot x)]\Gr\ra
  \no
  &= (\vepsilon_1\cdot \vepsilon_2) \la \Gr, (k_1\cdot x)^2 \Gr\ra.
  \end{align}
  Thus $c_1=0=c_2$. 
  
  The expansion coefficient of the mixed term $d(k_1, \lambda_1;k_2, \lambda_2)$ is given by
  \begin{align}
  -(\vepsilon_1\cdot \vepsilon_2) \la \Gr, (k_1\cdot x) (k_2\cdot x) \Gr\ra+ 2 \la \Gr, (\vepsilon_1\cdot x) H (k_1\cdot x) H^{-1} (k_2\cdot x) H(\vepsilon_2\cdot x)\Gr\ra.
  \end{align}
  Using (\ref{VGr1}) and that $\tilde{H}$ and $\TGr$ can be chosen to be real,  we have
  \begin{align}
  &\la \Gr, (\vepsilon_1\cdot x) H (k_1\cdot x) H^{-1} (k_2\cdot x)H (\vepsilon_2\cdot x)\Gr\ra+(1\leftrightarrow 2)
  \no
  &=\frac{1}{2}\Big\{
  -\im \la \Gr, (\vepsilon_1\cdot x)(k_1\cdot x) (k_2\cdot x) (\vepsilon_2\cdot \tilde{p})\Gr \ra
  \no
  &-\la \Gr,  (\vepsilon_1\cdot x)(k_1\cdot \tilde{p})H^{-1} (k_2\cdot x) (\vepsilon_2\cdot \tilde{p})\Gr \ra 
  \no
  & +\im \la \Gr, (\vepsilon_1\cdot \tilde{p})(k_1\cdot x) (k_2\cdot x) (\vepsilon_2\cdot x)\Gr \ra
  \no
  &+\la \Gr,  (\vepsilon_1\cdot \tilde{p})(k_1\cdot x)H^{-1} (k_2\cdot \tilde{p})  (\vepsilon_2\cdot x) \Gr \ra +(1 \leftrightarrow 2)
   \Big\}
   \no
   &= \frac{1}{2} \Big\{
   \im \la \Gr, (k_1\cdot x) \big(
   (\vepsilon_1\cdot \tilde{p})(\vepsilon_2\cdot x)-(\vepsilon_1\cdot x)(\vepsilon_2\cdot \tilde{p})
   \big)(k_2\cdot x)\Gr\ra
   \no
   &-\la \Gr, (\vepsilon_1\cdot x) (k_1\cdot \tilde{p}) H^{-1} (k_2\cdot x)(\vepsilon_2\cdot \tilde{p})\Gr\ra
   \no
    &+\la \Gr, (\vepsilon_1\cdot \tilde{p}) (k_1\cdot x)
   H^{-1} (\vepsilon_2\cdot x) (k_2\cdot \tilde{p})\Gr\ra+(1 \leftrightarrow 2)
   \Big\}
   \no
   &= (\vepsilon_1\cdot \vepsilon_2) \la \Gr, (k_1\cdot x)(k_2\cdot x)\Gr\ra+(\vepsilon_1\cdot k_2)(\vepsilon_2\cdot k_1) (\tfrac{1}{3}\la \Gr, x^2 \Gr\ra)
   \no
   &= \tfrac{1}{3} \la \Gr, x^2\Gr\ra \big(
   (\vepsilon_1\cdot \vepsilon_2)(k_1\cdot k_2)+(\vepsilon_1\cdot k_2) (\vepsilon_2\cdot k_1)
   \big).
  \end{align}
  This proves the assertion. $\Box$

  \section{Numerical coefficients}\setcounter{equation}{0}
\label{NumC}
  In this appendix, we will explain how to compute the following integrals appearing in the main text:
  \begin{align}
  S_1=& \int \dm k_1 \dm k_2 \frac{|k_1| |k_2|}{|k_1|+|k_2|}\,  \ex^{\im (k_1+k_2)\cdot \n} (1+(\hat{k}_1\cdot \hat{k}_2)^2),\label{SS1}\\
  S_2=& \int \dm k_1 \dm k_2 \frac{|k_1||k_2|}{|k_1|+|k_2|}\,  \ex^{\im (k_1+k_2)\cdot \n} (\hat{k}_1\cdot \hat{k}_2)(1-(\hat{k}_1\cdot \hat{k}_2)^2),\label{SS2}\\
  S_3=& \int \dm k_1 \dm k_2 \frac{|k_1||k_2|}{|k_1|+|k_2|}\,  \ex^{\im (k_1+k_2)\cdot \n} (1-(\hat{k}_1\cdot \hat{k}_2)^2)^2.\label{SS3}
  \end{align}
  These integrals are of the form
    \begin{align}
  S_j=& \int \dm k_1 \dm k_2 \frac{|k_1||k_2|}{|k_1|+|k_2|}\,  \ex^{\im (k_1+k_2)\cdot \n} F_j(\hat{k}_1\cdot \hat{k}_2),
  \end{align}
 which we  rewrite as 
   \begin{align}
  S_j=&\int_0^{\infty} \dm t \int \dm k_1 \dm k_2  |k_1| |k_2| \ex^{-t(|k_1|+|k_2|)}\,  \ex^{\im (k_1+k_2)\cdot \n} F_j(\hat{k}_1\cdot \hat{k}_2).
  \end{align}
  
  Let us switch  to  polar coordinates $(r,  \vphi, \vartheta)$ by
  \begin{align}
  \hat{k}&=(Y \cos \vphi, Y \sin \vphi, X),
  \no
  X&=\cos \vartheta,\ \ Y=\sin \vartheta.
  \end{align}
  Clearly $X^2+Y^2=1$. Then we have
  \begin{align}
  \hat{k}_1\cdot \hat{k}_2= \cos(\vphi_1-\vphi_2) Y_1 Y_2+X_1 X_2,
  \end{align}
  and hence 
  \begin{align}
  S_j
  =&\int_0^{\infty}\dm t \int_0^{\infty} \dm r_1 r_1^3 \ex^{-t r_1}  \int_{-1}^1\dm X_1\, \ex^{\im r_1 X_1}
  \no
  &\times \int_0^{\infty} \dm r_2 r_2^3 \ex^{-t r_2} \int_{-1}^1\dm X_2\, \ex^{\im r_2 X_2} 
    \mathfrak{S}_j (X_1, X_2),
  \end{align}
  where 
  \begin{align}
  \mathfrak{S}_j(X_1, X_2)&=\int_0^{2\pi}\dm \vphi_1 \int_0^{2\pi}\dm \vphi_2 F_j\big( \cos(\vphi_1-\vphi_2)Y_1Y_2+X_1X_2\big)
  \no
  &=2\pi \int_0^{2\pi} \dm \vphi F_j(\cos \vphi Y_1 Y_2+X_1X_2). \label{VphiInt}
  \end{align}
  Set 
  \begin{align}
  \la A\ra=2\pi \int_0^{\infty} \dm r\,  r^3\, \ex^{-t r} \int_{-1}^1 \dm X\, \ex^{\im r X}A(X).
  \end{align}
  After performing $\vphi$-integration in (\ref{VphiInt}), $S_j$ can be expressed as 
  \begin{align}
  S_1=&\int_0^{\infty} \dm t\Big\{
  \tfrac{3}{2} \la 1\ra \la 1\ra-\la 1\ra\la X^2\ra+\tfrac{3}{2}\la X^2\ra \la X^2\ra
  \Big\},\label{S1}
\\
S_2=&\int_0^{\infty} \dm t\Big\{
  -\tfrac{1}{2} \la X\ra \la X\ra+ 3 \la X\ra\la X^3\ra-\tfrac{5}{2}\la X^3\ra \la X^3\ra
  \Big\},  \label{S2}
  \\
  S_3=&\int_0^{\infty} \dm t\Big\{
  \tfrac{3}{8} \la 1\ra \la 1 \ra+ \tfrac{1}{2} \la 1\ra\la X^2\ra+\tfrac{3}{2}\la X^2\ra \la X^2\ra
  +\tfrac{3}{4}\la 1\ra \la X^4\ra-\tfrac{15}{2} \la X^2\ra \la X^2\ra
  \no
  &+\tfrac{35}{8} \la X^4\ra \la X^4\ra
  \Big\}.\label{S3}
    \end{align}
  Using  Mathematica, one obtains
  \begin{align}
  \la 1\ra&= \frac{-4+12t}{(1+t^2)^3},\\
  \la X\ra&=  \im \frac{16 t }{(1+t^2)^3},\\
  \la X^2\ra&= \frac{4(-3 +t^2)}{(1+t^2)^3},\\
  \la X^3\ra&=4\im \Big\{
   \frac{t(9+8t^2+3t^4)}{(1+t^2)^3}-3\, \mathrm{arccot}( t)
   \Big\}
   ,
   \\
   \la X^4\ra&=
   \frac{4(3+27 t^2+32 t^4+12 t^6)}{(1+t^2)^3}-48 t\,  \mathrm{arccot}( t)
   .
  \end{align}
  Inserting these formulas to (\ref{S1})-(\ref{S3}) and using  Mathematica again, one arrives at
  \begin{align}
  S_1 =92 \pi^3,\ \ S_2=208 \pi^3,\ \ S_3=256 \pi^3.
  \end{align}

\end{document}